\documentclass[twocolumn,showpacs,preprintnumbers,amsmath,amssymb,prb,aps]{revtex4-1}
\usepackage{epsfig}
\usepackage{graphicx}
\usepackage{dcolumn}
\usepackage{bm}
\usepackage{tabularx}

\begin{document}


\title{Complex hybridization physics and evidence of structural anomaly to be a bulk property in an exotic Fe-based compound, CaFe$_2$As$_2$}

\author{Ram Prakash Pandeya}
\author{Arindam Pramanik}
\author{Anup Pradhan Sakhya}
\author{A. Thamizhavel}
\author{Kalobaran Maiti}
\altaffiliation{Corresponding author: kbmaiti@tifr.res.in}

\affiliation{Department of Condensed Matter Physics and Materials Science, Tata Institute of Fundamental Research, Homi Bhabha Road, Colaba, Mumbai - 400 005, INDIA.}

\date{\today}

\begin{abstract}
Surface of quantum materials often exhibits significantly different behavior than the bulk due to changed topologies and symmetry protections. The outstanding problem is to find out if the exoticity of a material is linked to the changed topology at the surface or it is a bulk property. Hard $x$-ray photoemission spectroscopy (HAXPES) is a significantly bulk sensitive technique (escape depth of valence electrons is about 40 \AA\ for 6 keV photon energy) and the probing depth can be tuned by changing the electron emission angle. Therefore, HAXPES is often used to reveal the surface-bulk differences in a material. Here, we show that the delineation of surface-bulk differences in the valence band spectral functions using this method is highly non-trivial due to the complexity arising from linear dichroic effect in addition to the change in surface sensitivity. We show that core level spectra can be used to reveal the surface-bulk differences in the electronic structure. The Ca 2$p$ spectra exhibit evidence of significant hybridization with the conduction electrons revealing their importance in the electronic properties of the system as also found for the charge reservoir layers in cuprate superconductors. The Fe 2$p$ core level spectra as a function of bulk sensitivity and temperature reveals an unusual scenario; while the surface electronic structure corroborates well with the observed phase transitions of the system, the bulk spectra exhibit signature of additional structural phases providing a rare evidence of structural anomaly to be a bulk property.
\end{abstract}

\pacs{73.90.+f, 68.35.bd, 71.45.Gm, 75.30.Fv, 79.60.Bm}
\maketitle


Study of high temperature superconductors continues to attract tremendous attention due to various unresolved longstanding puzzles in these materials important for both fundamental physics and technological applications. Most of these materials forms in layered structure separated by an insulating layer, which is often called charge reservoir layer.\cite{chargereservoir} It is believed that these layers play an important role to derive superconductivity via preserving two dimensionality of the system and protecting the conduction layers via screening various types of disorder, which is introduced to achieve suitable charge carrier density. Recently discovered Fe-based superconductors also form in effective two-dimensional structure with intermittent insulating layer and exhibit additional complexity in the physics of these systems. For example, CaFe$_2$As$_2$, a 122-type Fe-based material, forms in tetragonal structure at room temperature and undergoes a transition\cite{Ni} to orthorhombic-antiferromagnetic phase below 170 K breaking the $C_4$ rotational symmetry to $C_2$-symmetry, which is often referred as nematicity. One can preserve the $C_4$ rotational symmetry by application of a small pressure ($>$0.35 GPa), which is called collapsed tetragonal (cT) phase although long range magnetic order gets destroyed.\cite{Kreyssig,Goldman,Pratt,Goldman2} Interestingly, a recent study found evidence of cT phase bands even within the ambient electronic structure.\cite{Khadiza-ARPES}

CaFe$_2$As$_2$ exhibits superconductivity at low temperatures under varied conditions such as application of  external pressure, chemical substitution at any of the three sites and quenching of the sample after/or during the preparation of the sample.\cite{Milton,Ca1-xRxFe2As2-Yanpeng,CaFe2-xIrxAs2-Yanpeng,CaFe2-xRhxAs2-Yanpeng,Saha,Kazutaka,Chen,Zhao} All these processes lead to the suppression of magnetic order. Pressure induced superconductivity is often accompanied by cT phase that led to controversy on the link between superconductivity and non-magnetic cT phase; while some studies support this idea, there are contrasting views indicating absence of superconductivity in the cT phase.\cite{Yu} It appears that lack of hydrostaticity of the applied pressure have an important role in the superconductivity of CaFe$_2$As$_2$. On the other hand, non-hydrostaticity in EuFe$_2$As$_2$ reduces the superconducting transition temperature.\cite{EuFe2As2-pressure}

Angle resolved photoemission spectroscopy (ARPES) studies of CaFe$_2$As$_2$ have been carried out using wide photon energy range (20-100 eV) to probe the near Fermi energy occupied electronic structure at different temperatures.\cite{KBM-Pramana,Ganesh-JAP,ARPES1} Since, the electron mean free path inside the material is around 10 \AA\ or lower in this photon energy range,\cite{surface} the ARPES measurements are highly surface sensitive. Therefore, it is not clear if anomalies observed in these materials are due to the complexity associated to the surface strain/reconstructions or it is a bulk phenomena. In order to address this outstanding puzzle, we employed hard $x$-ray photoemission spectroscopy (HAXPES) to study the electronic structure of CaFe$_2$As$_2$, which has significantly high bulk sensitivity.\cite{haxpes,SREP-Bi2Se3} We discover evidence of finite hybridization of the charge reservoir layer with the conduction states and unusual surface bulk differences involving the cT phase in the electronic structure.


High quality single crystals of CaFe$_2$As$_2$ were grown by high temperature solution growth method using Sn flux.\cite{crystal,crystal2} The crystal structure and chemical composition of the sample were confirmed using $x$-ray diffraction and energy dispersive analysis of $x$-rays. Magnetic susceptibility measurements exhibit sharp transition at 170 K indicating good quality of the sample. The HAXPES measurements were carried out at the P09 beamline at Pettra-III DESY, Hamburg using Phoibos electron analyser from Specs GmbH. Experimental setup was optimized to achieve the best energy resolution; 200 meV energy resolution was found at a photon energy of 5947.5 eV, which was used for all the measurements. In order to change the bulk sensitivity, measurements were carried out at different polar angles with respect to the surface normal. Fermi level was derived using valence band spectrum of Au mounted in the electrical contact with the sample. The sample was cleaved in ultrahigh vacuum condition and the photoemission measurements were done at a vacuum of 2$\times$10$^{-10}$ Torr. Electronic structure calculation was carried out using density functional theory (DFT) following the full potential linearized augmented plane wave method as captured in Wien2k software.\citep{wien2k} The convergence to the ground state was achieved self-consistently using 1000 $k$-points within the first Brillouin zone. We have used the Perdew-Burke-Ernzerhof generalized gradient approximation (PBE-GGA) for the density functional calculations.


\begin{figure}
\vspace{-2ex}
\includegraphics[scale = 0.45]{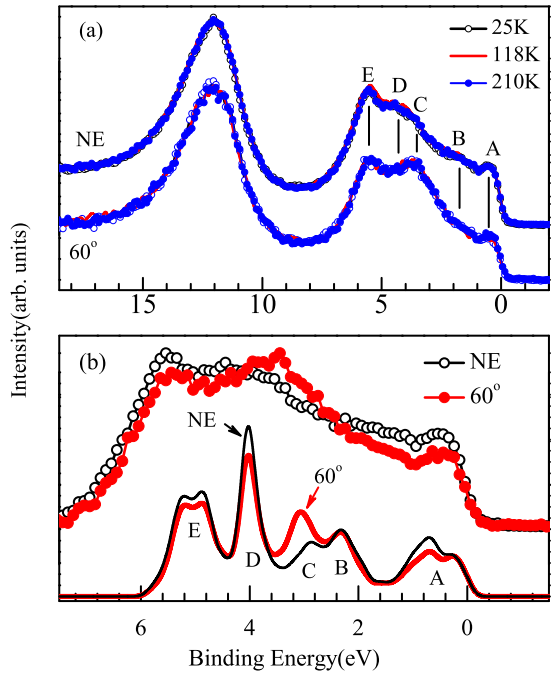}
\vspace{-16ex}
 \caption{(a) Valence band spectra at temperatures 210 K (solid circles), 118 K (solid line) and 25 K (open circles) at normal emission (NE) and 60$^o$ angled emission angles. (b) 210 K spectra at normal emission (open circles) and 60$^o$ emission (solid circles) are superimposed. The thin line (NE) and thick line (60$^o$ emission) represent the calculated spectral function after considering the cross section due to linear dichroic and transition matrix element effects.}
 \label{fig1:VB}
\end{figure}

In Fig. \ref{fig1:VB}, we show the valence band spectra exhibiting a feature around 12 eV binding energy corresponding to As 4$s$ photoemission signal and five distinct features denoted by A, B, C, D and E. The spectra collected at different temperatures (25 K - 210 K) at a particular experimental geometry overlap with each other almost perfectly although the system has undergone a concomitant structural and magnetic transition at 170 K. However, the relative intensities of the features at different emission angles are significantly different. The intensity pattern is very different from that in the Al $K\alpha$ spectra too.\cite{Ganesh-JAP} Difference with Al $K\alpha$ spectrum is complex due to the change in photoemission cross section in addition to the surface sensitivity. The drastic change in spectral intensities due to change in emission angle that correspond to different surface sensitivity of the technique may arise due to the difference in surface and bulk electronic structures.

From the electronic structure calculations and various ARPES studies, it is well known that the valence band in the vicinity of the Fermi level is constituted primarily by Fe 3$d$ states. Contributions from As 4$p$ states is large in 3 - 4 eV binding energies relative to that in the other energy regimes. The photoelectron cross-section \cite{cross-section} of Fe 3$d$ states is almost 10 times of the cross-section of As 4$p$ states for the photon energy range 27 - 100 eV. In the hard $x$-ray regime (photon energy $\sim$ 6-8 keV), Fe 3$d$ cross-section reduces to about one tenth of As 4$p$ cross-section. Therefore, in addition to high bulk sensitivity, the HAXPES is better to reveal the properties of the As 4$p$ states that has been found important due to significant covalency of these materials.\cite{Ganesh-FeTeSe,Kotliar-NPhys}

The other concern is the linear polarization of the incident beam. A change in emission angle significantly changes the angle between orbital axis and the incident light polarization vector, thereby, changes the corresponding photoemission cross-section. In order to verify such dichroic effect in the photoemission spectra, we have calculated the electronic structure of CaFe$_2$As$_2$; these results are consistent with the published data.\cite{Khadiza-SREP} Photoemission cross-section of the partial density of states (PDOS) is calculated in the following way: (i) First, we calculated the polarization dependent matrix elements of each of the Fe 3$d$ and As 4$p$ orbitals for different emission angles. (ii) The results were multiplied by the corresponding PDOS. Fe 3$d$ and As 4$p$ orbital contributions are added separately. (iii) The Fe 3$d$ and As 4$p$ contributions were multiplied by the photoemission cross-section to include the radial integration part of the matrix cross section.\citep{cross-section} (iv) To compare with the experimental results, we have convoluted the calculated spectral functions with the Fermi-Dirac distribution function and Gaussian (FWHM = 200 meV) representing resolution broadening. The final results are shown in Fig. \ref{fig1:VB}(b).

The simulated spectra at two different emission angles following this simple method reproduce{\bf s} the experimental scenario remarkably well. This suggests that the changes observed in the valence band spectra at different emission angles have large contribution from the linear dichroic effect in addition to the change in escape depth. The escape depth of the valence electrons in normal emission geometry of HAXPES ($h\nu \sim$ 6 keV) is close to 40 \AA\ and the bulk contribution in the normal emission spectra is about 86\% considering top 4 layers near the surface provide the upper limit of the surface layer thickness. At the emission angle of 60$^o$, the escape depth becomes 20 \AA, which is still quite large providing a bulk sensitivity of $\sim$74\%. All these observations suggest that it is difficult to probe the surface-bulk differences in the valence band of this material simply by changing the emission angle due to polarization induced effects. Change in photon energy is even more complex for such covalent system\cite{Ganesh-FeTeSe,Kotliar-NPhys} as it modifies the relative photoemission cross section of various constituent valence states significantly.

\begin{figure}
\includegraphics[scale = 0.4]{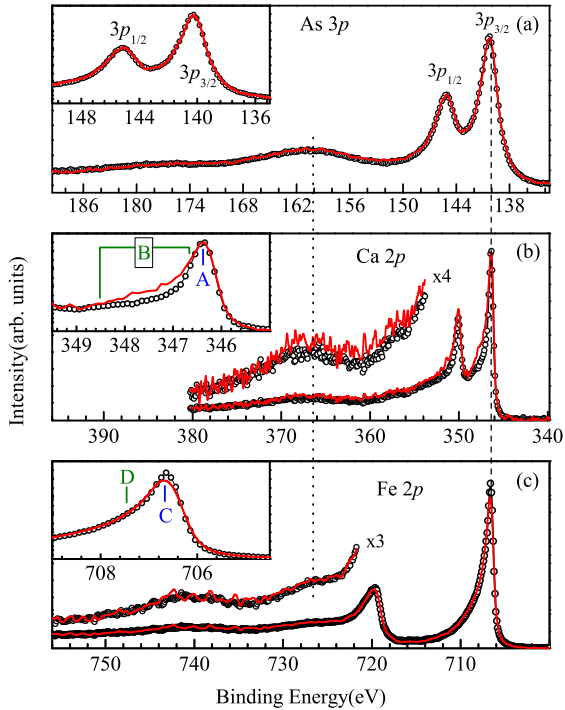}
\caption{Core level spectra of (a) As 3$p$, (b) Ca 2$p$ and (c) Fe 2$p$ collected at normal emission (open circles) and 60$^o$ angled emission (solid line) at 210 K. Weak intensities at higher binding energies are shown in renormalized intensity scale as mentioned in the figure. The insets show the same spectra in an enlarged energy scale.}
 \label{fig2:plasmon}
\end{figure}

Thus, we look for other method to find the evidence of differences between the surface and bulk electronic structures. It is well known that the photoemission spectra of the core levels are also derived by the eigenstates of the final state Hamiltonian containing interaction between the photo-hole and the valence electrons. In Fig. \ref{fig2:plasmon}, we show the core level spectra of As 3$p$, Ca 2$p$ and Fe 2$p$ collected at 210 K and two different electron emission geometries. Distinct features for the spin-orbit split peaks are observed in all the spectra exhibiting significant asymmetry due to excitation of electrons across the Fermi level along with the photoexcitation of the core electrons. This suggests metallic ground state of this material, which is consistent with the theoretical\cite{Khadiza-SREP} and experimental studies.\cite{Khadiza-ARPES,Khadiza-epjb}
There are intense additional broad features around 20 eV away from the  As 3\textit{p} core level peaks. Similar energy difference of these features from the main peak in every core level spectra indicate that these are related to the energy loss due to collective excitations such as plasmon excitations.\cite{SREP-Bi2Se3} The experimental spectra collected at different emission angles are shown by superimposing over each other and the highest angular momentum peak is shown in the inset. The plasmon peaks do not show significant angle dependence. The As 3$p$ spectra shown in Fig. \ref{fig2:plasmon}(a) is found to be insensitive to the change in probing depth of the technique.

Ca 2$p$ spectra shown in Fig. \ref{fig2:plasmon}(b), exhibit similar intensities in the higher binding energy region (355-380 eV). The intensity of the shoulder region of the main peak [e.g. $\sim$ 347.5 eV for 2$p_{3/2}$ signal shown by `B' in the inset of Fig. \ref{fig2:plasmon}(b)] is enhanced in the 60$^o$ emission case relative to the intensity of feature `A' suggesting its link to the surface electronic structure. CaFe$_2$As$_2$ is known to cleave at the Ca-layer keeping about 50\% of the Ca atoms on each of the cleaved surfaces. Such surface in the 122-class of materials often show significant reconstruction.\cite{SurfReconstruct} Thus, the electronic property of the surface Ca is expected to be different from the bulk Ca. In the Fe 2$p$ data, the higher binding energy regime (725-760 eV) appears quite similar for both the emission angles. However, the main peaks exhibit an interesting scenario [see the inset in Fig. \ref{fig2:plasmon}(c)]; the intensity at the peak position, `C' diminishes in the surface sensitive case while the intensity in the remaining part of the spectra remains unchanged.

\begin{figure}
\includegraphics[scale = 0.45]{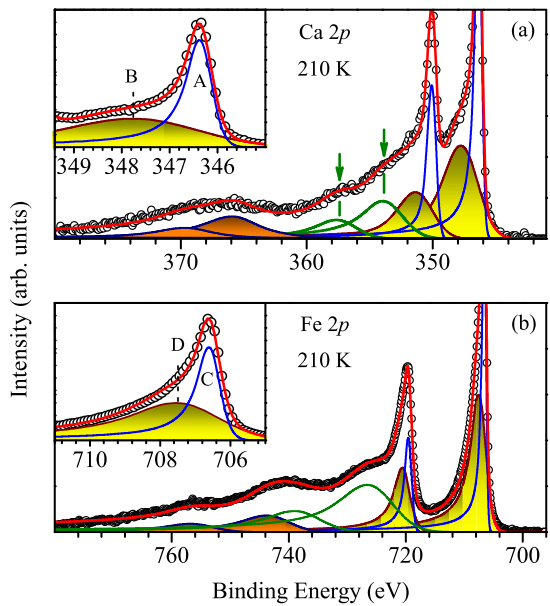}
\vspace{-20ex}
 \caption{(a) Ca 2$p$ and (b) Fe 2$p$ core level spectra. Open circles are the experimental data and superimposed lines are the simulated data. The constituent peaks are shown separately by line and area plot. Insets shows the higher angular momentum peak in an enlarged energy scale.}
 \label{fig3:Fit}
\end{figure}

In order to explore the underlying physics of these scenarios, we have simulated the experimental spectra using asymmetric Gaussian-Lorentzian product functions within the least square error method; the results are shown in Fig. \ref{fig3:Fit}. Asymmetric Gaussian-Lorentzian product functions contain effects due to energy resolution (Gaussian) and lifetime broadenings (Lorentzian). Plasmon peaks could be captured remarkably well considering two peaks with energy separation and intensity ratio identical to the spin-orbit split features in each case. The simulation of each of the spin-orbit split features in the Ca 2$p$ spectrum required at least two peaks; the features shown by yellow shaded area plot represent the shoulder intensities at 347.6 eV and 351.5 eV for Ca 2$p_{3/2}$ and 2$p_{1/2}$ signals, respectively. In the inset of Fig. \ref{fig3:Fit}(a), the scenario of 2$p_{3/2}$ is shown with two peaks at about 346.4 eV (Peak A) and 347.6$\pm$0.2 eV (Peak B). The experiment with higher surface sensitivity exhibit enhancement of Peak B intensity suggesting it to be a surface feature. The energy of the peak B is very similar to the Ca 2$p$ signal in CaO indicating effective divalency of surface Ca. The lower binding energy of the bulk Ca indicates its valency smaller than (+2) and hence, Ca 4$s$ band is not entirely empty due to hybridization induced effects.

In addition to these features, another set of features were required to simulate the intensities between 350-360 eV binding energies; signature of these features could also be seen in the experimental data (see arrows in the figure). This indicates that there are multiple final states associated to the Ca 2$p$ core level excitations in addition to the plasmon peaks. Such features in the core level spectra appear due to final state effects involving transfer of charge to screen the core hole.\cite{ruth-PRB} The presence of multiple features is an evidence of finite hybridization of Ca states with the ligand states that allows charge transfer for the core hole screening. Empty Ca 3$d$ levels appear far above the Fermi level making it energy expensive for charge transfer to Ca 3$d$ bands for the core-hole screening. Electronic structure calculations\cite{Khadiza-SREP} exhibit finite hybridization between Ca4$s$-As4$p$ states, which provides a pathway to charge transfer for core-hole screening.

The simulation of Fe 2$p$ spectrum shown in Fig. \ref{fig3:Fit}(b) also exhibit signature of multiple features. The depletion of intensity at 706.6 eV (peak C) [see the inset of Fig. \ref{fig2:plasmon}(c)] with the increase in surface sensitivity of the technique suggests that 706.6 eV peak corresponds to the bulk Fe 2$p$ signal while the surface feature appears at 707.7$\pm$0.2 eV (peak D). The higher binding energy of the surface feature may be attributed to the enhancement of surface potential arising from the changes in the surface Ca layer. From the analysis shown in Fig. \ref{fig2:plasmon}, it is clear that the feature around 727 eV binding energy is related to the loss due to plasmon excitation associated to the 2$p_{3/2}$ emission and the feature related to 2$p_{1/2}$ emission appears at 742 eV. The peaks at 740 eV and 754 eV may also be attributed to the plasmon excitations; in this case, the energy separation from the main peaks is double of the separation of the other plasmon peaks. It is noted here that Fe 2$p$ core level excitations often show satellite peaks\cite{DD,Fujimori} due to final state effects arising from the interaction between the valence electrons and the interaction of valence electron with the core hole. In the present case, it appears that one can simulate the Fe 2$p$ core level spectra without considering such features. It is not clear if such features are indeed absent or their distinct signature could not be probed in the experimental spectra due to overlapping broad plasmon peaks in the relevant energy regime.

\begin{figure}
\vspace{-2ex}
\includegraphics[scale = 0.45]{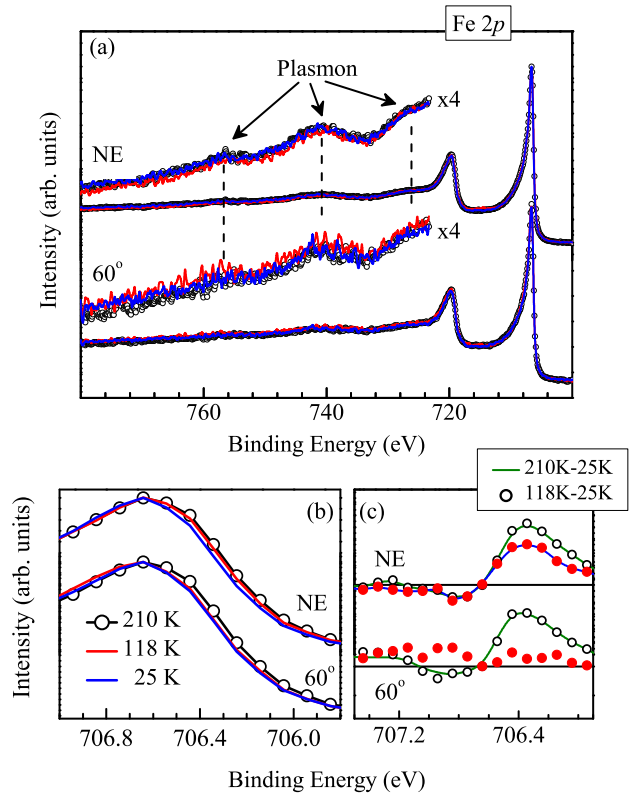}
\vspace{-8ex}
 \caption{(a) Fe 2$p$ core level spectra collected at normal emission and 60$^o$ angled emission at different temperatures. Weak intensities at higher binding energies are shown in an enlarged intensity scale as mentioned in the figure. (b) Fe 2$p_{3/2}$ core level region at different temperatures; 210 K (open circles), 118 K (red line) and 25 K (blue line) in an enlarged energy scale showing enhancement of width with the increase in temperature. (c) Difference spectra: (210K-25K) is shown by open circles and (118K-25K) by closed circles. Lines are hand drawn smooth curves.}
\label{fig4:Temp}
\end{figure}

In order to investigate the influence of structural and magnetic phase transition on the electronic structure, we have collected data at varied temperatures such as 210 K, 118 K and 25 K. Since, the valence band, in particular, near Fermi energy states are dominated by the Fe 3\emph{d} contributions and primarily derive the electronic properties, we focus on the evolution of the Fe 2\emph{p} core level spectra with temperature; As 3$p$ spectra (not shown here) do not show noticeable change with the change in temperature. Fe 2$p$ spectra shown in Fig. \ref{fig4:Temp}(a) exhibit identical lineshape and intensities of the features at higher binding energies (energy $\geq$ 722 eV) at different temperatures. Interestingly, the main peak exhibit a finite change with temperature, which can be interpreted as a small shift in energy or change in linewidth. The spectral modification is clearly visible as shown in Fig. \ref{fig4:Temp}(b); significant change despite large intrinsic width of the features. The width of the 60$^o$ emission spectra is same at 118 K and 25 K, and becomes larger above the transition temperature of 170 K. This is evident in the difference spectra shown in Fig. \ref{fig4:Temp}(c); (118K-25K) data exhibit almost no change while there is significant intensity at about 706.3 eV in the (210K-25K) data. Curiously, the normal emission spectra exhibit gradual increase in intensity at the same energy in both the cases as demonstrated in the subtracted data in Fig. \ref{fig4:Temp}(c).

It is well established that the valence electrons are involved in screening the core hole - electrons from ligand levels hops to the unoccupied part of the local electronic structure of the photoemission site. Thus, the width of the valence band will get convoluted resulting into additional width of the screened core level peaks. The larger width at 210 K observed here can be attributed to the larger valence bandwidth in the tetragonal structure possessing $C_4$ symmetry. Below the structural phase transition, the system evolves to orthorhombic structure having $C_2$-symmetry (nematic phase). Consequently, the bandwidth reduces - shrinking of Fermi surface at lower temperatures due to structural transition has indeed been observed in earlier studies.\cite{Khadiza-SREP} The narrowing of the width at lower temperatures may be attributed to such changes in the electronic structure. While this scenario is consistent with the evolution of the surface spectra, the scenario of bulk electronic structure appears anomalous.

Recent ARPES study\cite{Khadiza-ARPES} discovered signature of the energy bands related to the cT phase even in ambient conditions of the valence bands. The Fermi surface remain almost unchanged down to about 100 K. At lower temperatures ($<$ 100 K), one of the three hole pockets around the $\Gamma$-point vanishes and the overall band structure assume effective three dimensionality.\cite{Khadiza-ARPES,2dto3d} Thus, the low temperature width is expected to be narrower due to the band structure effect. The presence of larger width below the structural transition temperature of 170 K can be attributed to the presence of cT phase in the bulk possessing $C_4$-symmetry indicating that the presence of cT phase within the orthorhombic structure is a bulk property.


In conclusion, we studied the electronic structure of CaFe$_2$As$_2$ using hard $x$-ray photoemission spectroscopy. Although the surface sensitivity of the technique could be changed by changing photoelectron emission angle, the experimental valence band spectra exhibit complexity due to the linear dichroic effect that makes it difficult to delineate the surface-bulk differences of the valence band. The core level spectra, however, is less influenced by such effect and we observe significantly different surface and bulk features in the Ca 2$p$ and Fe 2$p$ core level spectra. The surface peaks appear at higher binding energies, which is attributed to the surface potential created by 50\% occupancy of the Ca-layer on the cleaved surface. The Ca 2$p$ spectra exhibit signature of satellites indicating significant Ca 4$s$-As 4$p$ hybridization. The temperature dependence of the Fe 2$p$ linewidth reveals unusual scenario. While surface is usually expected to be anomalous due to defects, surface reconstruction, symmetry protection, etc., the surface electronic structure of CaFe$_2$As$_2$ found here follows observed structural/magnetic transition and the hidden structural phases appears to be a bulk property.

Authors acknowledge the financial support under DST-DESY program and thank Dr. Andrei Gloskovskii for his support during the experiments. KM acknowledges financial assistance from the Department of Science and Technology, Government of India under J. C. Bose Fellowship program and the Department of Atomic Energy under the DAE-SRC-OI Award program.

\end{document}